\documentstyle[preprint,aps]{revtex}
\begin{document}
\draft
\title{Scheme of the replica symmetry breaking for short- ranged Ising
spin glass within the Bethe- Peierls method.}
\author{\rm K. Walasek}
\address{ Institute of Physics, 
Pedagogical University, Plac
S{\l}owia\'nski 6,\\ 65- 069 Zielona G\'ora, Poland.} 
\maketitle
\begin{abstract}
Within the Bethe- Peierls method the for short- ranged Ising spin glass
,recently formulated by Serva and Paladin, the equation for the spin
glass parameter function near the transition to the paramagnetic
phase has been carried out. The form of this equation is
qualitatively similar to that for Sherrington- Kirpatrick model, but
 quantitatively  the order parametr function depends
of the dimension $d$ of the system. In the case $d\rightarrow\infty$
one obtains well known Parisi solution.   
\end{abstract} 
The study of spin glasses (SG)'s in finite dimensions is very active since it is still
unclear if they share some the qualitative features of the mean-
field theory of the model Sherrington- Kirkpatrick (SK)\cite{1a,1}
However, there are recent investigations
\cite{2,3a,3,4} which indicate difficulties to extend the 
mean field approximation (MFA) scenario to realistic spin glasses with 
short- range
interaction and decide "a priori" which properties survive and which
must be appropriately modified.

The one of the first attempt to go beyond MFA was an
expansion of the SG order parameter for $d$ -dimensional hypercubic
lattice in $1/d$ \cite{5}. It has turned out that in this case, qualitatively, 
the Parisi's ansatz \cite{1} holds.  
 Recently, in an interesting paper \cite{6}, an approach
beyond the MFA has been achieved for an $d$- dimensional Ising SG
model with short- range interactions on a real lattice using an
extension of the Bethe- Peierls approximation (BPA) to the
spin glass problem via the replica trick. This approach seems to be
very promising to estabilish a direct contact with the results
obtained by different authors for the infinite- ranged version and to
controll possible deviations for short- ranged glasses from the well
acquired MFA scenario.  Quite recently \cite{7} the Parisi's scheme has
been investigated for the Ising SG with $S= 1/2$ using the
generalized of the Bethe- Peierls method named by the
authors "a variational approach" where finite clusters of spins
interacts and the sample averaging is properly taken into account.
The result for the free energy is qualitatively similar to that obtained in 
the frame of the MFA with some quantitatively modifications due to short- 
range order interactions. In particular in ref. \onlinecite{7} the two
dimensional system is studied numerically on the one step replica
symmetry breaking (RSB).  

The aim of this Letter is to show that using only the Bethe- Peierls
ansatz it is possible to obtain in an explicit form near the SG
transition the form of the SG order parameter for all stages RSB. As
usual the Hamiltonian of our system reads:
\begin{equation} \label{1} H=-\frac{1}{2}\Sigma_{i,j} J_{i,j}\sigma_i\sigma_j
\;\;,\end{equation} where $\sigma_i=\pm 1$ and $J_{i,j}$ are random variables
obeying dichotomic distribution, that is $J_{i,j}=\pm J$ with the
equal probability for $+$ and $-$ sign. Using the replica trick
combined with the PBA (see ref. \cite{6}) one obtains for the cluster
consisting of the central spin and $2d$ its nearest neighbours the
following effective Hamiltonian:\begin{equation}\label{2}
H_{\text{eff}}=\frac{-1}{\beta}\ln\left[\exp\left(\beta\sum_{i,\alpha}J_{0,i}
\sigma_{0,\alpha}\sigma_{i,\alpha}\right)\right]_{\text{av}}-\frac{\beta
J^2}{2}\sum_{\alpha\neq
\alpha^{\prime}}\sum_{i=1}^{2d}\mu_{\alpha,\alpha^{\prime}}\sigma_{i,\alpha}
\sigma_{i,\alpha^{\prime}}\;\;, \end{equation} where
$\left[\cdots\right]_{\text{av}}$ denotes sample averaging indices $0$ and
$i=1\cdots 2d$ refer to the central and lateral spins of the cluster,
respectively, $\alpha\neq \alpha^{\prime}=1\cdots n$ are replica indes
with $n\rightarrow 0$ at the end of calculations. In eq. (\ref{2}) $\mu_{\alpha,
\alpha^{\prime}}$'s describe the interaction betwen the "external world" and 
the lateral spins of the replicated cluster. Couplings
$\mu_{\alpha,\alpha^{\prime}}$ are
calculated from the following equations:
\begin{equation}\label{3}
\langle \sigma_{i,\alpha}\sigma_{i,\alpha^{\prime}}\rangle =\langle
\sigma_{0,\alpha}\sigma_{0,\alpha^{\prime}}\rangle\;\;, \end{equation}
 where \begin{equation}\label{4}
\langle\cdots\rangle=\frac{\text{Tr}\exp\left(-\beta
H_{\text{eff}}\right)\cdots}{\text{Tr}\exp\left(-\beta H_{\text{eff}}\right)}  
 \;\;.\end{equation}
It is easy to show the near below the SG transition $\mu_{\alpha,
\alpha^{\prime}}$ is a very small parameter and reach zero at the SG
transition point and in paramagnetic phase. Therefore one can expect
that the
structure of $\mu_{\alpha,\alpha^{\prime}}$'s  is the same as of the SG order 
parameters $q_{\alpha,\alpha^{\prime}}=\langle \sigma_\alpha\sigma_{\alpha
^{\prime}}\rangle $. Keeping this in mind, in order to recognize the
form of the SG order parameter one expands the left and right hand 
side of (\ref{3}) into $\mu_{\alpha,\alpha^{\prime}}$ to the third order which
is relevant to the study of the structure of
$\mu_{\alpha,\alpha^{\prime}}$ near the critical point. After a
tedious but strighforward algebra we get the following equation:
\begin{equation}\label{5}
\left(2\tau\mu_{\alpha,\alpha^{\prime}}+A_d\sum_{\alpha_1}\mu_{\alpha,\alpha_1}
\mu_{\alpha_1,\alpha^{\prime}}+\frac{2}{3}B_d\mu_{\alpha,\alpha^{\prime}}^3\right)=0\;\;,
\end{equation} where $\tau=\left(T_c-T\right)/T_c$, $\beta_c=1/T_c$
with $T_c$ being the SG critical temperature (for the value of $T_c$
see ref \cite{6}),
$A_d=\frac{\beta_cJ}{\left(2d-1\right)^{1/2}}$ and $B_d=\frac{2\beta_c^3
J^3d}{\left(2d-1\right)^{3/2}}$.

Now one introduces the Parisi scheme changing
$\mu_{\alpha,\alpha^{\prime}}$ to $\mu\left(x\right)$ with $0<x<1$
\cite{1}. After this the eq. (\ref{5}) takes the form:
\begin{equation}\label{6}
2\tau\mu\left(x\right)-2A_d\mu\left(x\right)\int_0^1 \mu\left(y\right)dy-
A_d\int_0^x\left[\mu\left(x\right)-\mu\left(y\right)\right]^2+\frac{2}{3}
B_d\mu\left(x\right)^3=0\;\;\end{equation}
After the standard procedure \cite{1} one finds that
\begin{equation}\label{7} \mu\left(x\right)=\frac{A_d}{2B_d}x \;\;, 
\end{equation}for $0<x<x_1$. For $x_1\leq x\leq 1$ the parametr $\mu\left(x
\right)$ reach a plateau and its value is
\begin{equation}\label{8} \mu\left(x\right)=\frac{A_d}{2B_d}x_1\;\;\end{equation}
with \begin{equation} x_1=\frac{4\beta_cJd}{\left(2d-1\right)^{1/2}}\tau\;\;.
\end{equation}

It is easy to show that for the infinite dimension
$\left(d\rightarrow\infty\right)$, after the aprioprate rescaling \cite{6,8} $J\rightarrow
J\sqrt{2d}$ and changing $\mu_{\alpha,\alpha^{\prime}}$ to
$2q_{\alpha,\alpha^{\prime}}$ one obtains the well known
Parisi scheme for the Sherrington- Kirpatrick model. 

Therefore one concludes that in the frame of the BPA the Parisi
ansatz holds for a real $d$-dimensional hypercubic lattices. If one
assumes that the BPA aprroximation for the Ising SG (\ref{1}) on the Bethe 
lattice is exact \cite{9} one can infer that the Parisi's scheme is
valid for that case.

The author acknowledge the financial support of the Polish Committee for 
Scientific Research (K. B. N.), Grant No 2 P03B 034 11 .
    
\end{document}